\documentclass[11 pt]{article}%
\usepackage{amsmath}%
\usepackage{amsfonts}%
\usepackage{amssymb}%
\usepackage{epsf}%
\usepackage{hyperref}%

\def\singlespace {\smallskipamount=3.75pt plus1pt minus1pt
                  \medskipamount=7.5pt plus2pt minus2pt
                  \bigskipamount=15pt plus4pt minus4pt
                  \normalbaselineskip=15pt plus0pt minus0pt
                  \normallineskip=1pt
                  \normallineskiplimit=0pt
                  \jot=3.75pt
                  {\def\smallskip {\vskip\smallskipamount}}
                  {\def\medskip   {\vskip\medskipamount}}
                  {\def\bigskip   {\vskip\bigskipamount}}
                  {\setbox\strutbox=\hbox{\vrule
                    height10.5pt depth4.5pt width 0pt}}
                  \parskip 7.5pt
                  \normalbaselines}
\def\middlespace {\smallskipamount=5.625pt plus1.5pt minus1.5pt
                  \medskipamount=11.25pt plus3pt minus3pt
                  \bigskipamount=22.5pt plus6pt minus6pt
                  \normalbaselineskip=22.5pt plus0pt minus0pt
                  \normallineskip=1pt
                  \normallineskiplimit=0pt
                  \jot=5.625pt
                  {\def\smallskip {\vskip\smallskipamount}}
                  {\def\medskip   {\vskip\medskipamount}}
                  {\def\bigskip   {\vskip\bigskipamount}}
                  {\setbox\strutbox=\hbox{\vrule
                    height15.75pt depth6.75pt width 0pt}}
                  \parskip 11.25pt
                  \normalbaselines}
\def\doublespace {\smallskipamount=7.5pt plus2pt minus2pt
                  \medskipamount=15pt plus4pt minus4pt
                  \bigskipamount=30pt plus8pt minus8pt
                  \normalbaselineskip=30pt plus0pt minus0pt
                  \normallineskip=2pt
                  \normallineskiplimit=0pt
                  \jot=7.5pt
                  {\def\smallskip {\vskip\smallskipamount}}
                  {\def\medskip   {\vskip\medskipamount}}
                  {\def\bigskip   {\vskip\bigskipamount}}
                  {\setbox\strutbox=\hbox{\vrule
                    height21.0pt depth9.0pt width 0pt}}
                  \parskip 15.0pt
                  \normalbaselines}

\begin{document}
\begin{center}
{\bf {\Large Gravitational Collapse of an Infinite Dust Cylinder}}
\bigskip

{\large Sashideep Gutti$^{a}$\footnote{e-mail address: sashideep@tifr.res.in},
T. P. Singh$^{a}$\footnote{e-mail address: tpsingh@tifr.res.in},
Pranesh A. Sundararajan$^{b}$,
and Cenalo Vaz$^{c}$\footnote{e-mail address: vaz@physics.uc.edu}
}
\bigskip

{\it $^{a}$Tata Institute of Fundamental Research,}\\
{\it Homi Bhabha Road, Mumbai 400 005, India}
\medskip

{\it $^{b}$Birla Institute of Technology and Science,}\\
{\it Pilani 333031, India}

\medskip

{\it $^{c}$Physics Department, University of Cincinnati}\\
{\it Cincinnati, Ohio, USA}
\end{center}
\bigskip
\bigskip

\begin{abstract}
\noindent We examine the gravitational collapse of an infinite
cylindrical distribution of time like dust. In order to simplify the
calculation we make an assumption that the axial and azimuthal
metric functions are equal. It is shown that the resulting solution
describes homogeneous collapse. We show that
the interior metric can be matched to a time dependent exterior. We
also discuss the nature of the singularity in the matter region 
and show that it is covered.
\end{abstract}
\singlespace

\section{Introduction}

While there have been many analytical studies of spherical gravitational
collapse, there have been only few investigations of non-spherical
collapse, the obvious reason being the complexity of Einstein equations.
Thus most of our present understanding of cosmic censorship and the nature
of singularities stems from studies of spherical collapse. Given the impact
that the results of these studies have had on our understanding of the end
states of classical collapse, it is of interest to ask what may emerge from
studies of gravitational collapse with different topologies.

A very useful study of the collapse of an infinite dust cylinder
was initiated by Thorne \cite{tho}. This is described by the metric
\begin{equation}
ds^{2}=e^{(G-P)}\left(dt^{2}-dr^{2}\right)-e^{P}dz^{2}
-\alpha^{2}e^{-P}d\phi^{2}.
\label{tho}
\end{equation}
The metric functions depend on $t$ and $r$. This form of the metric can be
inferred by first writing down the static Weyl axisymmetric metric for an
infinite line Newtonian source
\begin{equation}
ds^{2}=e^{P}dt^{2} - \alpha^{2}e^{-P}d\phi^{2} -
e^{(G-P)}\left(dr^{2}+dz^{2}\right)
\label{wey}
\end{equation}
in which the metric functions depend only on $r$ and $z$. The transformation
$t\rightarrow iz$, $z\rightarrow it$ in this metric then yields the metric
(\ref{tho}). It is invariant under Lorentz transformations in the $(t,r)$
plane and can be used to represent the collapse of a cylindrical dust shell. Apostolatos and Thorne \cite{ato}
used the above metric to show that rotation can halt the collapse
of the cylinder. The formation of a naked singularity for the infinite
cylindrical shell collapse was established by Echiverria \cite{ech}.
Naked singularities for counter-rotating dust shells were shown to occur,
by  Goncalves and Jhingan  \cite{jhi} and by Nolan \cite{nol}, and an exact
dynamical solution was given by Pereira and Wang \cite{pwa}.
The collapse of infinite cylindrical dust clouds has been studied also
by Chiba \cite{chi}, by Senovilla and Vera \cite {sen} and by 
Bondi \cite{bondi}.

Although the collapse of an infinite cylinder may not
constitute a ``realistic'' collapse scenario, it can simulate the collapse of
a finite ``bar'' or spindle-like matter distribution very near the central
regions of the spindle. In this sense the collapse of an infinite cylinder
is of astrophysical interest. More importantly, however, the problem probes
the structure of the general theory of relativity. Our analysis in the present
paper is of an infinite cylindrical cloud of time-like dust and is motivated
by what is known for spherical time-like dust collapse, {\it i.e.,} the
Tolman-Bondi spacetime. We make one assumption, namely that the azimuthal and
axial metric components are equal to each other. As we show, this assumption
implies that the solution is homogeneous. Also, the resulting system is a 2-d
system in which the metric depends on a radial coordinate, $r$ and on time,
$t$.

\section{Collapse of an infinite dust cylinder}

Consider an infinite cylindrical cloud of pressure-less, inhomogeneous,
time-like dust described by the stress tensor
\begin{equation}
T_{\mu\nu} = \epsilon(x)u_\mu u_\nu,
\end{equation}
where $u^2 = -1$. We first set up the metric for the interior of this infinite
dust cylinder, using comoving coordinates $(t,r,z,\phi )$. Assuming that $z$
represents the axis of symmetry of the cylinder, the metric functions depend
only on $t$ and $r$. The dust cloud is assumed to extend up to some radius
$r_{s}$; matching to a vacuum exterior will be discussed in Section 5. 
As a consequence of axisymmetry the functions $g_{t\phi }$,
$g_{r\phi }$ and $g_{z\phi }$ all vanish. Moreover, for an infinite cylinder,
invariance under $z\rightarrow -z$ implies that $g_{tz}$ and $g_{rz}$ also
vanish. We can use the freedom of coordinate transformations from the pair
$(t,r)$ to a new pair $(t^{\prime }, r^{\prime })$ to set two functions to
zero: these are $g_{tr}$ and the radial velocity $u^r$. As a result of the
latter condition, $r$ is determined to be a comoving coordinate. Since the
matter is dust, the metric can be comoving and synchronous, so that $g_{00}=-1$.
The metric for an infinite dust cylinder can hence be written in terms of
three unknown functions of two variables as
\begin{equation}
\label{met}ds^2=-dt^2+L^2(t,r)dr^2+M^2(t,r)dz^2+B^2(t,r)d\phi ^2.
\end{equation}
The energy-density $\epsilon (t,r)$ being the only non-zero component of the
energy-momentum tensor, Einstein's equations for this spacetime are (dot and
prime denote derivatives w.r.t. $t$ and $r$, respectively):
\begin{equation}
\label{Grr}G_{rr}=0 : \frac {\ddot M}{M}+\frac {\ddot B}{B}+ \frac {\dot
M}{M}\frac {\dot B}{B}=\frac{B^{\prime }M^{\prime }}{BML^2},
\end{equation}
\begin{equation}
\label{Gzz}G_{zz}=0 : \frac {\ddot B}{B}+\frac {\ddot L}{L}+ \frac {\dot
B}{B}\frac {\dot L}{L}=\frac{B^{\prime \prime }L-B^{\prime }L^{\prime }}{BL^3%
},
\end{equation}
\begin{equation}
\label{Gpp}G_{\phi \phi }=0 : \frac {\ddot M}{M}+\frac {\ddot L}{L}+ \frac
{\dot M}{M}\frac {\dot L}{L}=\frac{M^{\prime \prime }L-M^{\prime }L^{\prime }%
}{ML^3},
\end{equation}
\begin{equation}
\label{Gtr}G_{tr}=0 : \frac {\dot M^{\prime}}{M}+\frac {\dot B^{\prime}}{B}=
\frac {\dot L}{L}\left( \frac{M^{\prime }}M+\frac{B^{\prime }}B\right) ,
\end{equation}
\begin{equation}
\label{Gtt}G_{tt}=\epsilon(t,r) : -\frac{B^{\prime \prime }L-B^{\prime }L^{\prime }%
}{BL^3}-\frac{M^{\prime \prime }L-M^{\prime }L^{\prime }}{ML^3}-\frac{%
B^{\prime }M^{\prime }}{BML^2}+\frac {\dot B}{B}\frac {\dot L}{L}+ \frac
{\dot M}{M}\frac {\dot L}{L}+\frac {\dot B}{B}\frac {\dot M}{M}=\epsilon.
\end{equation}
The conservation of the energy-momentum tensor gives the relation
\begin{equation}
\label{con}\epsilon (t,r)=\frac{2\psi (r)}{LMB}
\end{equation}
where $\psi (r)$ is an integration function, to be determined by the initial
data. We also note that adding equations (\ref{Grr}), (\ref{Gzz}) and (\ref
{Gpp}), and using (\ref{Gtt}) and (\ref{con}) gives the relation
\begin{equation}
\label{add}\frac {\ddot M}{M}+\frac {\ddot B}{B}+\frac {\ddot L}{L}=-\frac{%
\psi (r)}{LMB}.
\end{equation}
Instead of working with the five Einstein equations (\ref{Grr})-(\ref{Gtt})
we will work with the equivalent set given by the equations (\ref{Grr}), (%
\ref{Gtr}), (\ref{con}), (\ref{add}) and the difference (\ref{Gzz})$-$(\ref{Gpp}).

We now assume that $B(t,r)\equiv rM(t,r)$. For a finite axisymmetric object
like a spheroid this would probably not be allowed by Einstein equations
but, as we see below, the equations are self-consistent when this assumption is
made for an infinite cylinder. The study of this system may be thought of as 
a prelude to examining the most general case, when $B$ and $M$ are not related.
The physical meaning of the assumption is that the object shrinks at the same
rate along the radial direction and the axis, so that its `prolateness' or
`oblateness' does not change with time. More importantly, we show that this 
assumption implies that the matter distribution is homogeneous. It is not the
most general homogeneous solution, but a special case.

With the above assumption the metric (\ref{met})
becomes
\begin{equation}
\label{nme}ds^2=-dt^2+L^2(t,r)dr^2+M^2(t,r)\left[ dz^2+r^2d\phi
^2\right]
\end{equation}
and Einstein's equations for this metric are as follows: eqn. (\ref{Grr}) becomes
\begin{equation}
\label{acc}\frac{2{\ddot B}}B+\frac{{\dot B}^2}{B^2}=\frac 1{L^2}\left(
\frac{B^{\prime 2}}{B^2}-\frac{B^{\prime }}{Br}\right) ,
\end{equation}
while eqn. (\ref{add}) reduces to
\begin{equation}
\label{acc2}\frac{2{\ddot B}}B+\frac{\ddot L}L=-\frac{\psi (r)}{rLM^{2}}
\end{equation}
These two are the dynamical equations, while the remaining three equations
are constraints. Eqn. (\ref{Gtr}) now becomes
\begin{equation}
\label{co1}\frac{2{\dot M}^{\prime }}M+\frac{\dot M}{rM}=\frac{\dot L}%
L\left( \frac{2M^{\prime }}M+\frac 1r\right)
\end{equation}
whereas the difference (\ref{Gzz})$-$(\ref{Gpp}) gives
\begin{equation}
\label{co2}\frac{2M^{\prime }}M=\frac{L^{\prime }}L
\end{equation}
and the conservation equation (\ref{con}) now is
\begin{equation}
\label{co3}\epsilon (t,r)=\frac{2\psi (r)}{rLM^2}.
\end{equation}
Eqns. (\ref{acc})-(\ref{co3}) are the Einstein equations for the metric (\ref
{nme}). Eqn. (\ref{co2}) can be solved to give
\begin{equation}
\label{ech}L(t,r)=h(t)M^2(t,r)
\end{equation}
where $h(t)$ is a arbitrary function of time. Eqn. (\ref{co1})
implies that
\begin{equation}
\label{Gin}\left( \sqrt{r}M\right) ^{\prime }=g(r)L(t,r)
\end{equation}
where $g(r)$ is an integration function. Using (\ref{Gin}) and (\ref{ech})
in the first dynamical equation (\ref {acc}) gives
\begin{equation}
\label{dy1}2{\ddot B}B+{\dot B}^2=rg^2(r)-\frac{r^2}{4h^2B^2}
\end{equation}
while the second dynamical equation becomes
\begin{equation}
\label{dy2}B{\ddot h}+4{\dot B}{\dot h}=-\frac{r^3\psi (r)}{B^3}+\frac{r^2}{2hB^3}-
\frac{2rg^2h}B.
\end{equation}

\section {Exact solution for a class of homogeneous dust collapse}

We choose the scaling $B(0,r)=r$ at time $t=0$, when collapse is
assumed to begin. This implies $M(0,r)=1$. 
We also choose $h(0)=1$, which from (\ref{ech}) implies $L(0,r)=1$. From
Eqn. (\ref{Gin}), written at $t=0$, we
get the relation $g(r)=1/2\sqrt{r}$. Then, solving  (\ref{Gin}) yields
\begin{equation}
\label{sg1} \frac{1}{M}=h(t) + K(t)\sqrt{r}.
\end{equation}

In this equation, $h(t), K(t)$ are functions of time only.
Now the differential equations (\ref{dy1}) and (\ref{dy2}) should be
valid at all r (including $r=0$). Substituting $B=rM$ in equation 
(\ref{dy1}) gives 
\begin{equation}
 2{\ddot M} + {\dot
  M}^2 = \frac{1}{4r^{2}}-\frac{1}{4r^2h^2M^2}$$ $$= \frac{h^2-({h+K{\sqrt{r}}})^2}{4r^2h^2}
\end{equation}
The $1/r^2$ term in the denominator makes the equation invalid for
$r=0$. This can be avoided by setting $K(t) =0$, which implies $h=1/M$ and 
hence that $M$ is a function only of time.
 Similarly equation
(\ref{dy2}) gives 
\begin{equation}
\label{sg2}M{\ddot h}+4{\dot M}{\dot h}=-\frac{\psi (r)}{rM^3}+\frac{1}{2hr^2M^3}-
\frac{h}{2Mr^2}.
\end{equation}
The sum of the last two terms vanishes when we set $K(t)=0$. There is a 
$1/r$ in
the denominator of the coefficient of $\psi(r)$. Since the left hand side
is independent of $r$, we are constrained to choose
$\psi(r)=k_1r$, where $k_1$ is a constant.
Substituting for $\psi$ in equation (\ref{co3})  gives
\begin{equation}
\label{sg3}\epsilon(t,r)=\frac{2k_1}{hM^4}=2k_1h^3.
\end {equation}
Since $h$ is a function of time only, this means that the density distribution
is homogeneous. Explicit solution for the metric can be obtained by solving the
equation 
\begin{equation}
\label{sg4}
 2\ddot{M}M + \dot{M}^2=0
\end{equation}
which gives
\begin{equation}
\label{sg5}
M=(c_1t+c_2)^{2/3}
\end{equation}
Substituting the value of $M$ in equation (\ref{sg2}) gives the
relation between $k_1$ and $c_1$; $k_1=2c_1^2/3$. Also, with $h(0)=1$, and
assuming $c_1$ to be negative, we can write 
\begin{equation}
\label{sg6}M=c_1^{2/3}(t_0-t)^{2/3}
\end{equation} where
$t_0=1/|c_1|$. The metric can now be written as 
\begin{equation}
\label{sg7}ds^2=-dt^2 +k^{2}(t_0-t)^{4/3}(dr^2+dz^2+r^2d\phi^2)
\end{equation}
where $k$ is a constant. At time $t=t_0$ there occurs a curvature 
singularity. All the shells become
singular at the same time $t_0$. This is similar to the spherical 
Oppenheimer-Snyder dust collapse.

\section{Comparison with spherical dust collapse}

Let us compare this system of equations with those for spherical dust
collapse, described by the Tolman-Bondi spacetime:
\begin{equation}
\label{tb}ds^2=-dt^2+L^2(t,r)dr^2+R^2(t,r)\left[ d\theta ^2+\sin ^2\theta
d\phi ^2\right] .
\end{equation}
The Einstein equations for this metric are:
\begin{equation}
\label{sp1}G_{rr}=0:\frac{2{\ddot R}}R+\frac{{\dot R}^2}{R^2}=\frac
1{R^2}\left( \frac{R^{\prime 2}}{L^2}-1\right) ,
\end{equation}
\begin{equation}
\label{sp2}G_{\theta \theta }=G_{\phi \phi }=0:\frac {\ddot L}L+\frac {\ddot
R} R+\frac {\dot L}L\frac {\dot R}R=\frac{R^{\prime \prime }L-R^{\prime
}L^{\prime }}{RL^3},
\end{equation}
\begin{equation}
\label{sp3}G_{tr}=0:\frac{{\dot R}^{\prime }}R=\frac {\dot L}L,
\end{equation}
\begin{equation}
\label{sp4}G_{tt}=\epsilon :-\frac{R^{\prime \prime }L-R^{\prime }L^{\prime }%
}{RL^3}-\frac{R^{\prime 2}}{2L^2R^2}+\frac{{\dot R}^2}{2R^2}+\frac {\dot L}L
\frac {\dot R}R+\frac 1{2R^2}=\epsilon
\end{equation}
The conservation equation is
\begin{equation}
\label{sp5}\epsilon (t,r)=\frac{\psi (r)}{LR^2}.
\end{equation}
We also note that adding equations (\ref{sp1}) and (\ref{sp2}) and using (%
\ref{sp4}) gives
\begin{equation}
\label{sp6}\frac{2{\ddot R}}R+\frac {\ddot L}L=-\frac{\psi (r)}{LR^{2}}.
\end{equation}

These equations should be compared with those for the cylinder: Eqn. (\ref
{sp1}) should be compared with Eqn. (\ref{acc}), Eqn. (\ref{sp2}) should be
compared with Eqn. (\ref{Gzz}), Eqn. (\ref{sp3}) should be compared with
Eqn. (\ref{co1}), Eqn. (\ref{sp4}) should be compared with Eqn. (\ref{Gtt}),
Eqn. (\ref{sp5}) should be compared with Eqn. (\ref{co3}), and Eqn. (\ref
{sp6}) should be compared with Eqn. (\ref{acc2})

Thus, although there is similarity in the forms of the metrics (\ref{nme}) and
(\ref{tb}), the cylindrical metric (\ref{nme}) describes homogeneous collapse,
whereas the spherical metric (\ref{tb}) describes general inhomogeneous
collapse. This is because of the additional Einstein equation (\ref{co2})
which is not there in the spherical case.

\section{Matching the interior cylindrical metric to a static exterior}
The metric (\ref{sg7}) is valid in the interior of a collapsing
dust cloud. The exterior is vacuum. The above metric will be a solution of
Einstein equations if it can be successfully matched to a static or a
time dependent exterior. It can be easily guessed that the interior
metric might not match to a static exterior, since the quadrupole
tensor for the matter distribution is non-zero, and the collapsing system
emits gravitational waves. 

To verify explicitly the impossibility of static exterior vacuum solution, we assume a static exterior metric of the form
\begin{equation}
\label{sg9}ds^2=-e^AdT^2+e^BdR^2+e^CdZ^2+R^2d\Phi^2
\end{equation} where $A,B,C$ are functions of $R$.
Only the diagonal components of the  Einstein tensor survive.
\begin{equation}
\label{sg10}G_{00}=\frac{e^{A-B}(-2C'-RC'^2+B'(2+RC')-2RC'')}{4R}
\end{equation}
 \begin{equation}
\label{sg11}G_{11}=\frac{2C'+A'(2+RC')}{4R}
\end{equation}
\begin{equation}
\label{sg12}G_{22}=\frac{e^{C-B}(RA'^2-2B'+A'(2-RB')+2RA'')}{4R}
\end{equation}
\begin{equation}
\label{sg13}G_{33}=\frac{e^{-B}R^2(A'^2-B'C'+C'^2+A'(-B'+C')+2A''+2C'')}{4}
\end{equation}
All the components equal zero in vacuum. Solving the Einstein equations
gives
\begin{equation}
\label{sg14}C=l_1\ln\left(\frac{R}{R_0}\right) + l_2.
\end{equation}

\begin{equation}
\label{sg15}A=\frac{-2l_1\ln\left(\frac{R}{R_0}\right)}{2+l_1} + l_3.
\end{equation}

\begin{equation}
\label{sg16}B=\frac{l_1^2\ln\left(\frac{R}{R_0}\right)}{2+l_1} + l_4.
\end{equation}

$R_0$ is a length scale which, as seen below, gets fixed by the matching.
We can get rid of $l_2$ and $l_3$ by rescaling $T$ and $Z$. So there
are two unknown constants. (see also \cite{bondi}).

In the interior let the outermost shell be
labelled by $r=r_s$. The coefficient of $d\Phi^2$ is  interpreted
as the square of the radius. The expression $r_sk(t_0-t)^{2/3}$ represents the evolution of the radius of the outer boundary of the collapsing dust with 
time. Further, it can be assumed that $\Phi$ and $Z$ represent the same
coordinates in both interior and exterior, i.e. $\Phi=\phi$ and $Z=z$.
Consider the hypersurface $r=r_s$.
On this hypersurface, the interior metric is given by
\begin{equation}
\label{sg17}ds^2=-dt^2 +k^{2}(t_0-t)^{4/3}(dr^2+dz^2+r_s^2d\phi^2).
\end{equation}
 The exterior is given by (\ref{sg9}).

 The metric coefficients of $\phi$
and $z$ should be same in both the interior and exterior.
Equating the coefficient of $d\phi^2$ we get 
\begin{equation}
\label{sg18}R_s^{2}=r_s^2k^{2}(t_0-t)^{4/3}
\end{equation}
The subscript $s$ in $R_s$ is to indicate the hypersurface in the
exterior coordinates.
Similarly equating the coefficients of $dZ^2$ for both interior and
exterior gives
\begin{equation}
\label{sg19}e^{l_1\ln{R_s/R_0}}=k^{2}(t_0-t)^{4/3}
\end{equation}
which implies $l_1=2$ and $R_0=r_s$.
 So the exterior metric on the hypersurface is 
\begin{equation}
\label{sg20}ds^2=-\frac{r_s}{R_s}dT^2 +\frac{cR_s}{r_s}dR^2+
\left(\frac{R_s}{r_s}\right)^2dZ^2+r_s^2d\phi^2
\end{equation}
where $c$ is a constant to be determined by matching the second
fundamental form. Matching the remaining components of the metric on
the hypersurface yields
\begin{equation}
\label{sg21}-dt^2=-\frac{r_s}{R_s}dT^2+\frac{cR_s}{r_s}dR^2.
\end{equation}
This can be used to obtain the relation between the interior and exterior
time.
The second fundamental form \cite{misner} is given by
\begin{equation}
\label{sg24}\Phi=(-n_{\mu;\nu}dx^{\mu}dx^{\nu})
\end{equation} 
and the extrinsic curvature is given by
\begin{equation}
\label{sg26}K_{\mu\nu}=n_{\mu;\nu}
\end{equation}
where $n_{\mu}$ is the unit normal to the hypersurface.
For the interior metric given by (\ref{sg17}) the hypersurface is given
by $r=r_s$. Calculating $K_{\phi}^{\phi}$ yields
 \begin{equation}
\label{sg27}K_{\phi}^{\phi}=\frac{1}{R_s}
\end{equation}
 where $R_s$ is given by (\ref{sg18}).
Similarly for the exterior metric $K_{\phi}^{\phi}$ can be computed.
The normal to the hypersurface $R_s$ can be obtained by
differentiating (\ref{sg18}),

 \begin{equation}
\label{sg28}dR+\frac{(2r_s\sqrt{k}(t_0-t)^{-1/3}\frac{dt}{dT})dT}{3}=0.
\end{equation}
Using the above equation and evaluating $K_\phi^\phi$ gives 
 \begin{equation}
\label{sg29}K_{\phi}^{\phi}=\frac{n^R}{R_s}
\end{equation}
where $n^R$ is the contravariant component in the $R$ direction.
By equating $K_\phi^\phi$ for both interior and exterior we
 obtain the value for $c$,
 \begin{equation}
\label{sg33}c=\frac{1}{\frac{R_s}{r_s}-\frac{4r_s^{2}k^{3}}{9}}
\end{equation}
$c$ was initially assumed to be a constant. In equation (\ref{sg33})
 $c$ is dependent on $R_s$ which is given by (\ref{sg18}). So c is not
 a constant as assumed but is time dependent. Hence there is no static exterior for the metric
 given by (\ref{sg17}).

\section{Matching with time dependent exterior}
In this section it will be shown that the interior metric (\ref{sg17})
can be matched to a time dependent exterior.
Let us assume an exterior of the form,
\begin{equation}
ds^{2}=e^{(G-P)}(dT^{2}-dR^{2})-e^{P}dz^{2}
-\alpha^{2}e^{-P}d\phi^{2},
\label{tho1}
\end{equation}
 where $G,P,\alpha$ are functions of $R,T$. The Einstein equations are
\begin{equation}
\label{en1}G_{TT}=\frac{-\alpha(P'^2+\dot{P}^2)+2(G'\alpha'-2\alpha''+\dot{G}\dot{\alpha})}{4\alpha}=0
\end{equation}
\begin{equation}
\label{en2}G_{TR}=\frac{\alpha'\dot{G}-\alpha P'\dot{P}+G'\dot{\alpha}-2\dot{\alpha'} }{2\alpha}=0
\end{equation}
\begin{equation}
\label{en3}G_{RR}=\frac{-\alpha(P'^2+\dot{P}^2)+2(G'\alpha'-2\ddot{\alpha}+\dot{G}\dot{\alpha})}{4\alpha}=0
\end{equation}
\begin{equation}
\label{en4}G_{ZZ}=\frac{e^{-G+2P}(\alpha(P'^2+2G''-4P''-\dot{P}^2-2\ddot{G}+4\ddot{P})+4(-P'\alpha'+\alpha''+\dot{P}\dot{\alpha}-\ddot{\alpha}))}{4\alpha}=0
\end{equation}
\begin{equation}
\label{en5}G_{\phi\phi}=\frac{e^{-G}\alpha^2(P'^2+2G''-\dot{P}^2-2\ddot{G})}{4}=0
\end{equation}
Here prime and dot denote partial derivative with respect to $R$ and $T$
respectively.
In vacuum all the above expressions equal zero. Subtracting equations
(\ref{en1}) and (\ref{en3}) gives 
\begin{equation}
\label{en6}\alpha''-\ddot{\alpha}=0
\end{equation}
Equation (\ref{en5}) gives
\begin{equation}
\label{en7}G''-\ddot{G}=\frac{\dot{P}^2-P'^2}{2}
\end{equation}
Substituting equations (\ref{en6}) and (\ref{en7}) in equation
(\ref{en4}) gives
\begin{equation}
\label{en8}P''-\ddot{P}+\frac{P'\alpha'-\dot{P}\dot{\alpha}}{\alpha}=0
\end{equation} 
The remaining two equations can be written as 
\begin{equation}
\label{en300}\dot{G}\alpha'-P'\dot{P}\alpha+G'\dot{\alpha}-2\dot{\alpha}'=0
\end{equation}
\begin{equation}
\label{en310}-\alpha(P'^{2}+\dot{P}^{2})+2(G'\alpha'-2\alpha''+\dot{G}\dot{\alpha})=0.
\end{equation}
The equations (\ref{en6}), (\ref{en7}), (\ref{en8}) are second order
partial differential equations in $T$ and $R$.
The outer boundary of the collapsing matter (hypersurface $r=r_s$) is
a priori unknown in the exterior coordinates. It is of the form
$f_1(T,R)=const$.
To solve completely the partial differential equation (\ref{en6})
(second order in both $T$ and $R$), we need to know the value on the
hypersurface and also the first derivative in the direction perpendicular
to the hypersurface. We assume that $Z$ and $\Phi$ represent the
same coordinates in both the interior and exterior. So on the hypersurface the metric components have to
match.
Equating coefficient of $d\phi^2$ for both metrics we get
\begin{equation}
\label{en9}\frac{\alpha^2}{e^P}=r_s^2k^{2}(t_0-t)^{4/3}
\end{equation} 
Similarly equating coefficient of $dZ^2$ we get.
\begin{equation}
\label{en10}e^P=k^{2}(t_0-t)^{4/3}
\end{equation} 
Multiplying both we get
\begin{equation}
\label{en11}\alpha=k^{2}r_{s}(t_0-t)^{4/3}
\end{equation} 
To solve the equations of the type $\alpha''-\ddot{\alpha}=0$, we
already know the value on the hypersurface given by (\ref{en11}). We
need one more arbitrary function to provide information about the value of the derivative
in the perpendicular direction. Similarly we know the value of $P$ on
the hypersurface (\ref{en10}). To solve the equation (\ref{en8}) we
need one more arbitrary function. For solving equation (\ref{en7}) we
need two arbitrary functions since we do not know the value of $G$ on
the hypersurface. These four arbitrary functions along with the initial
hypersurface $f_1(T,R)$ makes five arbitrary functions.  The matching
will be possible if the five arbitrary functions are enough to satisfy
all the constraints which are required during matching.
The relation between the interior time and the exterior coordinates
can be obtained by the equation similar to equation (\ref{sg21})
\begin{equation}
\label{en12}-dt^2=e^{G-P}(-dT^2+dR^2)
\end{equation}
Equations (\ref{en300}) and (\ref{en310}) act as constraints and can be
used to eliminate two of the five arbitrary functions.
Now  matching the second fundamental form \cite{misner}
\begin{equation}
\label{en13}\Phi=K_{tt}dt^2 +K_{zz}dz^2+K_{\phi\phi}d\phi^2
\end{equation}
gives at most three independent constraints. There are three arbitrary
functions left which can take care of the constraints offered by the
second fundamental form.
So in principle the interior metric (\ref{sg17}) can be matched to the
time dependent exterior
 given by (\ref{tho1}).

\section{Nature of the singularity}

The question of interest is to find out whether the curvature singularity 
resulting in the matter interior from the gravitational collapse of the 
dust cloud is naked or covered. This question can be addressed by looking at 
the null geodesic expansion $\Theta$ for the radial null geodesics. The
non-zero components of the tangent to the null geodesic are
$K^{r}(t,r)$ and $K^{t}(t,r)$.  The calculation of $\Theta$ for the 
cylindrical metric (\ref{nme}) proceeds exactly as for the spherical case 
discussed in \cite{sing} and we get the
result
\begin{equation}
\label{nul}\theta =2K^r\frac{Z^{\prime }}Z
\left[ 1+\frac{\dot B}{\sqrt{r}g(r)}\right]=
2K^r\frac{Z^{\prime }}Z\left[ 1+2h(0)\dot B\right]=
\frac{K^{r}}{r}\left[1+2r\dot{M}\right]
\end{equation}
where $Z\equiv B/\sqrt{r}=\sqrt{r}M(t)$. $Z'/Z$ is positive,
because of equation (\ref{Gin}), and $K^{r}$ is positive because we are
examining outgoing geodesics.

$K^{r}$ and $K^{t}$ can be evaluated explicitly from the geodesic equations
\begin{equation}
\frac{dK^{t}}{dk} = - \frac{\dot{M}}{M} \left(K^{t}\right)^{2},\qquad
\frac{dK^{t}}{dk} = - 2\dot{M} \left(K^{r}\right)^{2}
\label{geo}
\end{equation}
Here, $k$ is the affine parameter and we have used the relation 
$dt/dr=M(t)=K^{t}/K^{r}$ for outgoing geodesics, which follows from 
using $ds^{2}=0$ for radial null geodesics. 

The explicit solution is however not necessary for our purpose. 
As the singularity is approached, i.e as $t$ tends to $t_0$, 
$\dot{M}$ tends to $-\infty$, and hence so does the geodesic expansion
$\theta$, showing that the singularity is covered. The expression for
$\theta$ is in fact identical to that for the spherical marginally bound
homogeneous dust solution [Oppenheimer-Snyder] for which the metric is
\begin{equation}
\label{ostb}ds^2=-dt^2+S^2(t,r)\left[dr^2+
 r^{2}d\theta ^2+r^{2}\sin ^2\theta
d\phi ^2\right] 
\end{equation}
and the scale factor $S(t)$ goes as $(t_{0}-t)^{2/3}$. Comparison of 
this metric with the cylindrical homogeneous metric (\ref{sg7}) 
makes it clear that the geodesic expansion $\theta$ behaves identically in 
the two cases.

It has earlier been demonstrated \cite{tho} that the collapse of an infinite 
dust
cylinder results in a naked singularity, in accord with the hoop conjecture.
How is our result above consistent with these earlier findings? A possible
explanation is that there is a singular Cauchy horizon in the exterior,
as discussed recently by Tod and Mena \cite{tm}.
It should be noted that our analysis here does not provide information about
the formation of a singularity, or otherwise, in the exterior spacetime.

\section{Conclusion}

In this paper we have considered the collapse of an infinite cylindrical 
cloud of time-like homogeneous and pressureless dust subject to the condition 
that the axial and azimuthal metric functions are equal. We have explicitly 
shown that the resulting solution is  homogeneous
and that the collapse terminates in a curvature
singularity. We have shown that there exists a vacuum exterior and it
can be demonstrated that the singularity forming in the matter region 
is covered.

\vskip 0.2 in

\noindent{\bf Acknowledgments}

It is a pleasure to thank Tomohiro Harada for helpful conversations, and
Sergio Goncalves for extensive and useful discussions and suggestions on an 
earlier version of the manuscript.

\end{document}